\documentclass[twocolumn,showpacs,preprintnumbers,amsmath,amssymb,floatfix]{revtex4}
\usepackage{graphicx}% Include figure files
\usepackage{dcolumn}% Align table columns on decimal point
\usepackage{bm}% bold math
\def\bp{\beta^{\prime}}
\newcommand{\cO}[0] {\mathcal{O}}

\begin{document}

\title{VASP on a GPU: application to exact-exchange calculations of the stability of elemental boron}

\author{M. Hutchinson}
\author{M. Widom}
\affiliation{
Department of Physics\\
Carnegie Mellon University\\
Pittsburgh, PA  15213
}

\date{\today}

\begin{abstract}
General purpose graphical processing units (GPU's) offer high processing speeds for certain classes of highly parallelizable computations, such as matrix operations and Fourier transforms, that lie at the heart of first-principles electronic structure  calculations.  Inclusion of exact-exchange increases the cost of density functional theory by orders of magnitude, motivating the use of GPU's.  Porting the widely used electronic density functional code VASP to run on a GPU results in a 5-20 fold performance boost of exact-exchange compared with a traditional CPU.  We analyze performance bottlenecks and discuss classes of problems that will benefit from the GPU.  As an illustration of the capabilities of this implementation, we calculate the lattice stability $\alpha$- and $\beta$-rhombohedral boron structures utilizing exact-exchange.  Our results confirm the energetic preference for symmetry-breaking partial occupation of the $\beta$-rhombohedral structure at low temperatures, but does not resolve the stability of $\alpha$ relative to $\beta$.
\end{abstract}

\maketitle

\section{\label{sec:Intro}Introduction}

First principles quantum mechanical calculations of total energy are among the most pervasive and demanding applications of supercomputers. The problem is an interacting many-body problem whose wavefunction depends on the coordinates of $N_e$ electrons.  The computational  complexity grows exponentially with the number of electrons~\cite{HeadGordon08}, thus severely limiting the number of atoms that can be treated.  Elemental boron, for example, with its highly complex crystal structure containing approximately 107 atoms, lies well beyond the limits of exact energy calculation.

Replacing the many-body problem with $N_e$ coupled single electron problems reduces the exponential dependence to a polynomial, at the cost of introducing approximations.  For example, both Hartree-Fock (HF) and electronic density functional theory (DFT) are formally $\cO(N_e^4)$ in complexity~\cite{HeadGordon96}.  However, further approximations can additionally reduce the dependence on $N_e$. Indeed, $\cO(N_e)$ methods are possible in principle provided the problem is sufficiently local~\cite{Goedecker99}.  In practice, for the physics problems to be discussed below (utilizing the plane-wave-based code VASP), the complexity of HF-type hybrid functionals~\cite{Paier05} is $\cO(N_e^3 \log N_e)$ and for DFT~\cite{Kresse96} it is $\cO(N_e^2 \log N_e)$.  The benefit of HF-type calculation is the exact treatment of electron exchange interactions, while the net cost amounts to orders of magnitude in run time.

While the run time could possibly be reduced by running on a faster computer, actual frequencies of computer chips have held rather constant in recent years.  Instead of running at higher frequencies, the trend has been to increase computer power through parallelization, using multi-core chips, multi-chip nodes, and multi-node computer systems.  Recently massively parallel processing became available for low-cost computer systems through the introduction of general purpose graphical processing units (GPU).  These systems can contain hundreds of cores, with a low cost and low power consumption per core.  It is thus of high interest to evaluate the suitability of GPU systems for practical electronic structure calculations.  The fast Fourier transformation (which is an important VASP bottleneck) has been ported to the GPU~\cite{Eck11}.  Here we address the hybrid HF-DFT functionals that include exact-exchange.

\begin{figure*}
\includegraphics[width=4.0in,angle=90]{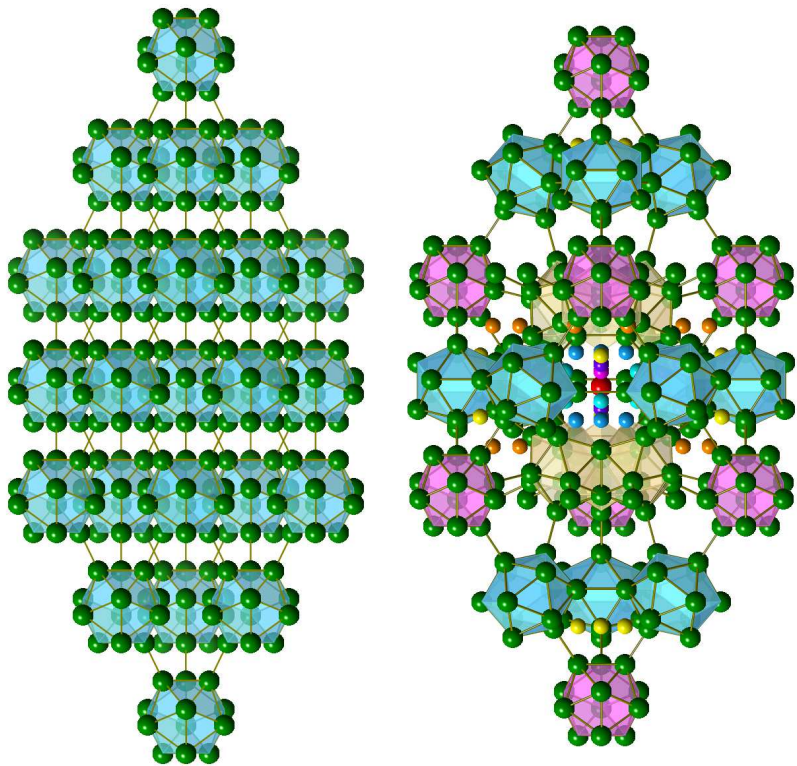}
\caption{\label{fig:alpha} (color online) (left) Structure of
  $\alpha$-rhombohedral Boron. (right) Structure of
  $\beta$-rhombohedral Boron.  Fully occupied sites shown in
  green. Fully occupied B15 site at cell center in red.  Partially
  occupied sites are: B13 (75\%) blue; B15 100\%; B16 (27\%) yellow;
  B17 (9\%) yellow; B18 ( 7\%) magenta; B19 (7\%) maroon; B20 (4\%)
  orange.
  %[WIDOM: colors are confusing (two yellows?  maroon, magenta, and orange are hard to distinguish.  there seem to be two different blues)]
  }
\end{figure*}

Below, we first first summarize the physics problem to be addressed, then describe our implementation of VASP on a GPU system, and finally we apply it to study the energies of competing boron structures.  Our key results are: 1) exact-exchange calculations confirm that the symmetric $\beta$ structure is energetically unfavorable relative to $\alpha$, but its energy can be reduced through symmetry breaking partial site occupancy.  2) The GPU system outperforms the CPU, with speedups reaching a factor of 20 in computationally demanding cases.

\section{\label{sec:boron}Elemental boron}

Boron is of practical interest because of its light weight, high strength and semiconducting properties.  It is also intrinsically interesting owing to its many competing complex structures, notably the $\alpha$- and $\beta$-rhombohedral structures characterized by differing arrangements of icosahedral clusters, as illustrated in Fig.~\ref{fig:alpha}.

We implemented exact-exchange calculations within VASP on a GPU system and applied the method to calculate total energies of elemental boron structures.  Two factors motivate this decision. 1) Competing structures of elemental boron are sufficiently close in energy that approximations within DFT might affect the identification of the true low temperature state.  Hence HF-type calculations including exact-exchange are warranted. 2) The structures are quite complex with the true ground state possibly containing 107 or more atoms per unit cell, usually with low symmetry.  Hence the HF calculations will be highly demanding and the GPU system may be beneficial.

\subsection{Structural stability}

Determining the stable structure of elemental boron remains problematic after more than a half century of effort.  A substantial number of polymorphs are known~\cite{Hoard67}, including at least two rhombohedral and two tetragonal forms as well as an orthorhombic high pressure form~\cite{Oganov09}.  All polymorphs are built around a common motif of 12 atom icosahedral clusters.  Many of the polymorphs appear to be kinetically trapped, or stabilized by impurities.

Our goal is to determine the minimum enthalpy structure in the limit of low temperature and pressure.  The two relevant structures are the $\alpha$-rhombohedral structure~\cite{McCarty58} (Pearson type hR12, with 12 atomic sites per primitive cell), which was initially believed~\cite{Hoard60} to be the stable form stable at low temperatures below T=1000K, and the more complex $\beta$-rhombohedral structure~\cite{Hughes63,Hoard70,Geist70} (Pearson type hR105, with 105 atomic sites per primitive cell) which was initially believed to be stable at high temperatures, from T=1500K up to melting. Although uncertainty remains concerning the role of impurities~\cite{Muller99}, current opinion suggests $\alpha$ is only metastable, and $\beta$ is the true equilibrium state for all temperatures, from melting down to T=0K~\cite{Hoard67}.

Total energy calculations initially challenged the stability of the $\beta$ phase.  Early calculations on isolated clusters using a  molecular orbital model~\cite{Takeda93} gave hints that $\alpha$ might be lower in energy than $\beta$.  This was later demonstrated for bulk crystalline structures (specifically, the $\alpha$-hR12 and $\beta$-hR105 structures) using DFT~\cite{Mihalk04} and has been since reconfirmed~\cite{Masago06,Setten07,Widom08}.  The energy difference between structures is quite small, and not well resolved within DFT, ranging from $\Delta E_{\beta\alpha}$=48 meV/atom in the local density approximation (LDA) to $\Delta E_{\beta\alpha}$=26 meV/atom in the PW91~\cite{PW91} generalized gradient approximation (GGA).

A possible explanation lies in inaccuracies of the assumed hR105 structure. In fact, experimentally observed density is consistent with 106 or 107 atoms in the rhombohedral cell. The most recent and high quality refinement~\cite{Slack88} reports 141 atomic sites per rhombohedral cell, many with partial occupancy.  The optimal arrangement identified using the PW91 GGA~\cite{Widom08} now places 107 atoms among these sites and reaches $\Delta E_{\bp\alpha}$=-0.86 meV/atom.  That is, it predicts the partially occupied $\bp$ structure is {\em lower} in energy than $\alpha$.  
Experimental evidence exists for structural anomalies at low temperatures based on photoabsorption~\cite{Werheit70,Werheit84}, internal friction~\cite{Tsaga70,Tsaga09} and heat capacity~\cite{Werheit11}.  These phenomena could be consistent with the predicted low temperature symmetry-breaking phase transition~\cite{Widom08,Widom09}.

However, the energy of the PW91-optimized structure still exceeds the energy of $\alpha$ by  $\Delta_{\bp\alpha}$=15 meV/atom using LDA.  Given the very small energy differences (per atom) among competing structures, it is not certain that DFT is of sufficient accuracy to correctly address the relative stabilities.  In fact, the situation is reminiscient of a frustrated system~\cite{Ogitsu10}.  To further test the reliability of our energy calculations requires climbing the ``ladder of density functionals''~\cite{Perdew01}.  The proposed sequence is LDA, GGA, meta-GGA, hybrid.  The GGA differs from LDA through the inclusion of electron density gradient terms.  Meta GGA then adds second derivatives.    Unfortunately, within the VASP code only the  PKZB~\cite{Perdew99} meta-GGA is presently available, and that is included in a non-self-consistent fashion~\cite{Hirschl02}.  Hybrid functionals include exact-exchange.  We choose the HSE06~\cite{HSE06} functional as the best currently available.

\subsection{Setup of DFT calculations}

Our study focuses on four specific structures of interest: the 12-atom $\alpha$-rhombohedral structure with Pearson type hR12; a 96-atom supercell of $\alpha$, doubled along each axis in order to match the lattice parameters of $\beta$, that we denote hR12x8; the ideal 105-atom $\beta$-rhombohedral structure with Pearson type hR105; and the symmetry-broken 107-atom $\bp$ variant that optimizes the GGA total energy that has Pearson type aP107.

Because of the approximate doubling of lattice constants between $\alpha$ and $\beta$, we double the linear density of $k$-point meshes for hR12 runs compared with the other structures.  Because of the differing symmetries of the structures the numbers of independent $k$-points (NKPT) differ among structures in addition.  Figures for a typical mesh density are given below in Table~\ref{tab:tests}.  Run times are linear in NKPT for conventional DFT and quadratic for HF-DFT.  Also important are the grids used to represent charge densities and wavefunctions in real space, with the number of grid points in a line (NGX) proportional to the corresponding lattice parameter.  The time complexity of FFTs grow with an $N\log{N}$ factor for each spatial dimension, or $N^3 \log N$ overall.  The values quoted correspond to the VASP setting ``PREC=Normal'', in which FFT's are performed without wrap-around errors.

\begin{table}
\begin{tabular}{l|c|rr}
Name& $k$-mesh & NKPT & NGX\\
\hline
hR12   & 4x4x4 & 10 & 24   \\
hR12x8 & 2x2x2 &  2 & 48   \\
hR105  & 2x2x2 &  2 & 48   \\
aP107  & 2x2x2 &  4 & 48   \\
\end{tabular}
\caption{\label{tab:tests} Parameters for typical runs.
  $k$-point meshes are Monkhorst-Pack. NKPT is the number of
  symmetry-independent $k$-points, and NGX the linear dimension of the
  FFT grid.}
\end{table}

We utilize the Projector Augmented Wave (PAW) potentials and adopt the PBE~\cite{PBE} generalized gradient approximation for our standard exchange correlation functional.  PBE is the basis on which the HF-type hybrid functional known as HSE06 is built~\cite{HSE06}.  Wavefunctions are represented in reciprocal space with a plane wave energy cutoff of 319 eV (the default setting). For future reference we show the convergence of results with respect to mesh density in Table~\ref{tab:conv}.  Evidently a 3x3x3 mesh size will be adequate for resolving energy differences at the meV/atom level that is needed for comparison of structural energies.  However, for computational efficiency, our benchmarking will focus on the 2x2x2 meshes discussed in Table~\ref{tab:tests}, as they demonstrate non-trivial meshes, but also small enough to be easily studied.

\begin{table}
\begin{tabular}{r|rr|rr}
  & \multicolumn{2}{c|}{PBE} & \multicolumn{2}{c}{HF} \\
\hline
$k$-mesh &$\Delta_{\beta\alpha}$&$\Delta_{\beta'\alpha}$&$\Delta_{\beta\alpha}$&$\Delta_{\beta'\alpha}$\\
\hline
1x1x1 &   -6.87&  -16.45&  +10.92&   -7.67\\
2x2x2 &  +24.39&   -1.06&  +44.25&   +7.11\\
3x3x3 &  +26.63&   -0.17&  +46.74&   +8.06\\
4x4x4 &  +26.07&   -0.20&        &        \\
\end{tabular}
\caption{\label{tab:conv} Table of $k$-mesh convergence. $k$-point meshes are Monkhorst-Pack. Units are meV/atom.  Energies labeled hR12 are for the super cell hR12x8.}
\end{table}

\section{\label{sec:porting}Porting of VASP}

VASP is widely used for DFT quantum chemistry calculations.  It is a large Fortran code, with contributions dating back to the 1980's, and portions of the code written in English, French and German.  VASP's performance has been discussed in the past~\cite{Paier05,Kresse96}, with the most relevant discussion by Eck et al~\cite{Eck11}.  The exact-exchange calculations involve less-commonly used functionality based on principles from HF theory.  However, this functionality consists of numerical operations similar to those used in more conventional calculations, so the performance enhancements obtained should transfer to more conventional portions of the code.

\subsection{CPU Performance Analysis}

\begin{table}
\begin{tabular}{l || r r r r}
Component    &    hR12 &   hR12x8 &    hR105 &    aP107 \\
\hline \hline
FOCK\_ACC    &   0.6208 &  8.6188 & 10.3627 & 21.2870 \\ 
FOCK\_FORCE  & 1.2776 & 16.7303 & 19.9418 & 41.0646 \\
Other        &  0.0235 &    0.1883 &    0.2087 &    0.3622 \\
\hline
Overall      &    53.0 &    730.2 &    877.1 &  1,802.2 \\
\end{tabular}
\caption{\label{tab:work}
Computational cost, measured in TFLOP, of mutually exclusive sections of VASP.  Tests are specified in Table~\ref{tab:tests} and run with a single electronic and ionic minimization step.  Overall costs are projected assuming a total of 5 ionic minimization steps and 75 electronic minimization steps.}
\end{table}

Before considering performance analysis, we define the test platform and test cases.  This study was performed on tirith (tr), a midrange workstation, and blacklight (bl), a SGI UV 1000 supercomputer at the Pittsburgh Supercomputing Center~\cite{PSC}.  Tirith is equipped with an Intel Core i7 920, which is a Nehalem-based quad-core clocked at 2.67 GHz with 8MB of cache, 12GiB of DDR3 at 1333 MHz, a Tesla C2075 GPU, and a Tesla C2050 GPU.  Tirith's current value is around \$7,000.  Tirith runs Intel Composer XE 2011, providing compilers and BLAS through MKL, FFTW version 3.2.2, and the 4.0 version of CUDA toolkit, which includes cuBLAS, cuFFT, and compilers.  Blacklight is a 256 blade system, each blade housing two Intel Xeon X7560s, which are Nehalem-based 8-core chips clocked at 2.27 GHz with 24 MiB cache, and 128 GiB DDR3 at 1066 MHz.  Blacklight runs Intel compilers and BLAS version 11.1 and FFTW version 3.2.2.  The Integrated Performance Monitor (IPM)~\cite{Skinner2005} runs on both systems and provides timing and hardware counter information which can be broken down across user-defined sections of the code. This allowed for the measure of floating-point operations.

Using the boron test structures in Table~\ref{tab:tests}, the efficiency of VASP on the CPU can be characterized.  The LINPACK benchmark~\cite{Linpack} provides an estimate of the peak realizable performance: 11.19 GFLOPs on a single CPU core.  Floating-point performance of VASP is found by dividing the workloads given in units of floating point operations (FLOP, or TFLOP using standard power-of-10 SI prefixes) in Table~\ref{tab:work} by the run-times given in units of seconds (s) in~\ref{tab:eff}, yielding units of FLOPs (FLOP/s).  In the test cases above, VASP runs at only 1.0-1.75 GFLOPs, or about 9-16\% utilization. 

The exact-exchange functionality is primarily contained in two routines: FOCK\_ACC and FOCK\_FORCE.  FOCK\_ACC applies the Fock exchange operator, while FOCK\_FORCE calculates the Hellman-Feynman forces acting on the ions.  The two routines are nearly identical, with a literal copy of FOCK\_ACC constituting the majority of the FOCK\_FORCE routine.  A set of FOCK\_ACC calls are made once per electronic minimization step, while a set of FOCK\_FORCE calls occur once per ionic minimization step.

IPM reveals the computational workload of these two routines.  For our structures, they account for over 98\% of floating-point operations and over 97\% of run-time, as seen in Table~\ref{tab:work}.  FOCK\_ACC requires approximately half as much effort as FOCK\_FORCE, per call, but is called much more frequently, making it the dominant routine.

Because FOCK\_ACC and FOCK\_FORCE constitute the overwhelming majority of each electronic and ionic minimization step, respectively, we can project the workload and run-time of full-length runs from data collected in truncated test runs.  The total time for a full run is
\begin{equation}
t_{\rm run}=S_e t_e + S_i t_i+{\rm Other}
\end{equation}
where $S_e$ and $S_i$ are, respectively, the numbers of electronic and ionic minimization steps. $t_e$ and $t_i$ are the corresponding time increments, and Other represents all the remaining portions of the code. This method is also used to project computational cost, where times are replaced by Floating-Point Operations in the formula above.  Note that this method underestimates by neglecting the non-FOCK, per-minimization operations, which, along with the Other category, have been shown to be minimal.

Within the FOCK routines, performance can be broken down even further.  On the CPU, BLAS calls comprise 20-30\% of the run-time and FFT calls comprise between 35-50\%.  The remaining 25-30\% is distributed between data manipulation kernels and `book-keeping'.  This includes particular effort spent in scatter/gather operations used to express cutoffs in the projected regions.  The heavy dependence on BLAS and FFT makes the VASP code a prime candidate for GPU acceleration.

\begin{table*}[t]
  \centering
  \begin{tabular}{l || rr | rr | rr | rr}
  Structure    & \multicolumn{2}{c|}{hR12} & \multicolumn{2}{c|}{hR12x8} & \multicolumn{2}{c|}{hR105} & \multicolumn{2}{c}{aP107} \\
  \hline
  Platform     & cpu      & gpu     & cpu       & gpu      & cpu       & gpu      & cpu         & gpu      \\
  \hline \hline
  FOCK\_ACC (s)   & 409.9 &   59.9 &  5,093.8 &    387.3 & 10,467.2 &    487.8 & 12,866.0 &    984.2 \\
  FOCK\_FORCE (s) & 789.1 & ~290.1 & 10,714.9 & ~1,199.3 & 22,144.5 & ~1,435.5 & 22,598.0 & ~2,912.8 \\
  Other (s)       &  26.9 &   27.5 &    117.8 &    134.6 &    216.2 &    142.2 &    206.6 &    246.3 \\
  \hline
  Overall (hr)    &  9.64 &  1.66 &   121.04 &    9.77 &   248.88 &   12.20 &   299.49 &   24.62 \\
  Speedup         & \multicolumn{2}{c|}{5.82x} & \multicolumn{2}{c|}{12.39x} & \multicolumn{2}{c|}{20.41x} & \multicolumn{2}{c}{12.17x} \\
  \end{tabular}
  \caption{\label{tab:eff}Run-times of components of VASP exact-exchange runs.  Units are seconds (s) and hours (hr).  Tests are specified in Table~\ref{tab:tests} and run with a single electronic and ionic minimization step.  Overall times are projected assuming a total of 5 ionic minimization steps and 75 electronic minimization steps.  CPU runs are single-core and GPU runs are single-device.}
\end{table*}

\subsection{Our Implementation}
Our implementation aims to be a proof of concept for GPU acceleration of exact-exchange in VASP.  We tried to make the implementation as simple and compartmental as possible, while preserving the abstraction and interfaces inherent in VASP.  At five points in the VASP code, our implementation intercepts the normal flow of execution.  The first two are trivial: in the main routine we intercept at the beginning and end to create and destroy library contexts and global variables.  The third intercepts the FFT3D routine that occur outside the FOCK routines, sending large FFTs to the GPU while leaving small ones on the CPU.  The final two intercepts bypass the FOCK\_ACC and FOCK\_FORCE routines discussed previously.

The optimal size at which to begin sending FFTs to the GPU can be computed once per system with a simple benchmark that directly compares CPU and GPU run-time for FFTs.  On our system, the optimal size was found to be $28^3$.  This decision does not apply to FFTs in the FOCK\_ACC and FOCK\_FORCE routines.  Recent improvements in the cuFFT library have reduced the importance of using power-of-two FFT grids compared to previous reports~\cite{Eck11}.

The two exact-exchange calculations consist of four nested loops, the outer two over two k-point indices and the inner two over band indices.  The loops compute several quantities for every pair of bands in the structure.
For example, the magnitude of the Fock exchange energy is
\begin{equation}
\frac{e^2}{2}\sum_{\substack{{\bf k}n\\{\bf q}m}} f_{{\bf k}n} f_{{\bf q}m}
\int d^3{\bf r}d^3{\bf r'}\frac
{\phi^*_{{\bf k}n}({\bf r})\phi^*_{{\bf q}m}({\bf r}')
\phi_{{\bf k}n}({\bf r}')\phi_{{\bf q}m}({\bf r})}
{|{\bf r}-{\bf r}'|}
\end{equation}
where the outer loops run over $k$-points ${\bf k}$ and ${\bf q}$ and the inner loops run over bands $m$ and $n$.  We place the inner loops over bands on the GPU.  The inner-most loop is multiplexed in a manner similar to that found in other parts of the CPU code and controlled by the NSIM parameter.  This changes matrix-vector operations into more efficient matrix-matrix operations, increasing the data size and allowing for the application of thousands of GPU threads.  It also allows the memory usage on the GPU, which is dependent on the number of bands processed concurrently, to be controlled at run-time.  By moving non-intensive routines to the GPU (i.e. every operation that occurs within the two loops), memory transfers between the host and device are minimized.  The port operates within the existing parallelism in VASP, distributing data across processors with MPI.

The GPU code relies heavily on the cuFFT and cuBLAS libraries for performing efficient FFT and BLAS operations, respectively.  An additional 20 custom kernels were written to replicate the CPU's functionality.  CUDA streams are used to define functional dependencies, allowing for asynchronous concurrent execution of smaller kernels, boosting performance on small input structures.  Two numerical precision settings are available: full double precision and mixed precision, which evaluates some or all FFTs in single and everything else in double.  The two precisions settings were found to agree within one thousandth of a percent, comparable to numerical differences seen when running the same structure on different platforms.

\begin{table*}
  \centering
  \begin{tabular}{l | c || r r r| r r r r }
  Structure & k & T-1C0G & T-1C1G & T-2C2G & B-16C & B-32C & B-64C & B-128C \\ 
  \hline 
  hR12   & 1 &      98.3 &     90.3 &     64.8 &     43.3 &     47.8 &    60.5 &   172.4 \\
  hR12x8 & 2 &  14,464.7 &  1,650.7 &    983.6 &  1,964.8 &  1,206.0 & 1,070.7 & 1,160.3 \\
  hR105  & 2 &  15,530.5 &  2,097.2 &  1,075.2 &  2,157.0 &  1,201.1 & 1,039.7 & 1,221.0 \\
  hR105  & 3 & 160,148.6 & 20,489.9 & 10,318.0 & 21,080.4 & 10,741.3 & 7,794.9 & 5,817.5 \\
  aP107  & 2 &  32,178.2 &  3,748.4 &  2,168.4 &  4,452.5 &  2,515.4 & 1,900.9 & 1,816.5 \\
  \end{tabular}
  \caption{Actual run-times of truncated runs, reduced NELM and NSW, of different structures on different platforms.  T is tirith, B is blacklight, attributes
$m$C$n$G indicates $m$ CPU cores and $n$ GPU devices.}
  \label{tab:times}
\end{table*}

\subsection{Performance Results}
There are a number of ways to compare CPU and GPU performance.  The simplest is to look at run-times.  Using IPM, we are able to accurately measure the time of the FOCK\_ACC and FOCK\_FORCE routines, along with the overall time.  For the boron structures in Table~\ref{tab:tests} and timings in Table~\ref{tab:eff}, we see that the GPU performs at 7x-21x in the FOCK\_ACC routine and at 3x-15x in the FOCK\_FORCE routine.  For large structures, a small speedup is also seen in the remaining, non-FOCK code, with the GPU running at up to 1.5x.  Assuming the computational cost is the same for the CPU and GPU version, the GPU utilization can be calculated.  NVIDIA's CUDA accelerated LINPACK running on a single GPU on our system performs at around 250 GFLOPs.  Our VASP implementation runs at around 9-26 GFLOPs, or about 4-10\% utilization.  

Another interesting comparison is the tirith workstation vs. the blacklight supercomputer.  We tested boron structures on the CPU and GPU of tirith, and 4 different CPU configurations on blacklight: 16, 32, 64, and 128 core.  As seen in Table~\ref{tab:times}, the gpu workstation performs comparably to 32-64 cores on the supercomputer for the structures of interest.

For small structures, blacklight becomes saturated with excess communication at a relatively low core-count, a well-known limitation of the scaling of VASP.  Tirith, however, behaves like a `fat' node in that the GPU's compute power is greatly increased but the inter-process communication stays fixed.  This leads the GPU to compare favorably to many-CPUs on small structures, as seen in the second row of Table~\ref{tab:times} labeled hR105 $k$=2, where we see that 2 CPU cores plus 2 GPUs perform equivalently to the fastest time on the supercomputer, which occurs for 64 cores.

There is, however,  a limit to this effect.  If the FFT grids are exceptionally small, then the single GPU performance is so poor that scaling effects don't make a difference, as seen in the first row of Table~\ref{tab:times}.

On larger structures, the GPU efficiency dramatically increases, causing it to compare favorably to runs on a small number of CPUs.  Large CPU runs benefit from the large system as well, though, making the comparison between the GPU and a large number of CPUs less favorable.

% won't need this when we have real values
Note that the discrepancies between Table~\ref{tab:times} and Table~\ref{tab:eff}, seen clearly in the hR105 case, are expected.  Table~\ref{tab:times} was constructed using truncated runs, increasing the relative impact of non-minimization overhead, which is negligible in the projections in Table~\ref{tab:eff}.

\section{Conclusions}

Our exact-exchange calculations confirm that the fully occupied hR105 structure of $\beta$-boron is energetically unfavorable with respect to $\alpha$, and that symmetry-breaking assignments of partially occupied sites of the hR141 structure substantially reduce the total energy per atom.  However, our specific realization of atomic positions within hR141 that was optimized using the PW91 functional (GGA) proves unstable (see Table~\ref{tab:nrgs}) relative to $\alpha$ in our exact-exchange calculation.  It appears that we need to re-optimize the structure of hR141 before we know the relative energies of $\alpha$ and $\beta$ within the exact-exchange framework.

\begin{table}
\begin{tabular}{l|rrrrr}
                &  LDA& PW91&  PBE& PKZB&   HF\\
\hline
$E_{\beta\alpha}$&47.83&25.87&26.63&37.02&46.74\\
$E_{\beta'\alpha}$&15.48&-0.86&-0.17& 8.53& 8.06\\
\end{tabular}
\caption{\label{tab:nrgs} Table of structural energies (units meV/atom). Here  $\beta$ refers to the ideal hR105 structure, $\beta'$ refers to the 107 atom optimized variant of B.hR141.  Energies of $\alpha$ are obtained from the super cell hR12x8.  All values are given for the 3x3x3 $k$-point mesh.}
\end{table}

The notion of a ladder of density functionals~\cite{Perdew01} implies the existence of a sequence of calculational methods that converge towards an exact answer, presumably one that confirms experimental reality.  Inspecting the series of energies in Table~\ref{tab:nrgs} we see that indeed the values appear to converge.  Surprisingly they converge towards values that are closer to the local density approximation than to the generalized gradient approximation.  Still, the remaining variability among density functional suggests that an even higher level of theory (e.g. quantum Monte Carlo) might be needed to fully resolve the problem of the low temperature stable crystal form of boron.

The GPU implementation of hybrid functionals in VASP outperformed one CPU core by about an order of magnitude, allowing the study of boron structures with exact-exchange.  It would take 64 CPU cores on the blacklight supercomputing to achieve the same time-to-solution for the hR105 test case with two k-points.  Even more remarkably, the minimal time-to-solution achievable on blacklight with any number of cores is only 3\% shorter than our 2 GPU system. This turns a supercomputing-level calculation into a tractable problem that can be addressed on-demand on desktop hardware.

Just as this speedup enables us to apply hybrid functionals to boron, it can enable other previously impractical exact-exchange calculations such as structural transitions in pnictide superconductors.  The previously high cost of entry for this type of calculation, namely a large cluster or supercomputer access, has slowed the application of exact-exchange DFT calculations to practically sized structures.  We have directly shown the feasibility of exact-exchange on moderately complex structures (hundreds of ions), and project that multi-node GPU clusters could be used to apply exact-exchange to even larger structures.

GPU acceleration could have positive effects on general DFT as well.  As mentioned previously and shown in Eck et al.~\cite{Eck11}, the methods used in this port are transferable to more conventional VASP use-cases.  VASP scales poorly across large numbers of nodes in large computer systems, so it can be optimal to treat large complex structures using a small number of `fat' nodes.  Previously, this meant spending lots of money on high-end chips, which would provide at most a factor of two in performance improvement (e.g. by going to high frequencies or wider SIMD instructions).  Now, nodes can be fattened with GPUs, providing order-of-magnitude improvements at low cost.  Since the communication overhead for a GPU and CPU are similar, it is reasonable to expect the largest GPU-based system to scale beyond the largest CPU-based systems by the same factor that a single GPU outperforms a single CPU.  Based on our analysis, that means a factor of 2 or three increase in structure size or, in the case of molecular dynamics, an order of magnitude increase in achievable simulation times.

\begin{acknowledgments}
This work was supported by the PETTT project PP-CCM-KY02-123-P3.  This research was supported in part by the National Science Foundation through TeraGrid resources provided by Pittsburgh Supercomputing Center.  We thank the NVIDIA corporation for use of their video cards and for useful discussions.  We thank Profs. Hafner, Kresse and Marsman for their advice, support and encouragement.
\end{acknowledgments}

\bibliography{gpu}

\end{document}